\documentstyle[twocolumn,prl,aps,epsf]{revtex}

\begin{document}
\draft

\title{Evidence for Spinodal Decomposition in Nuclear Multifragmentation}

\author{B.~Borderie$^1$, G.~T\u{a}b\u{a}caru$^{1,2}$, Ph.~Chomaz$^3$,
M. Colonna$^4$, A.~Guarnera$^4$, M.~P\^arlog$^2$, 
M.F Rivet$^1$  \\
   and  \\
 G.~Auger$^3$, Ch. O. Bacri$^1$, N.~Bellaize$^5$, R.~Bougault$^5$, 
 B.~Bouriquet$^3$, R.~Brou$^5$,
  P.~Buchet$^6$, A.~Chbihi$^3$, J.~Colin$^5$,
 A.~Demeyer$^7$,
   E.~Galichet$^{1,8}$,
  E.~Gerlic$^7$, D.~Guinet$^7$, S.~Hudan$^3$, P.~Lautesse$^7$, F.~Lavaud$^1$,
 J.L.~Laville$^3$, J.F.~Lecolley$^5$, C.~Leduc$^7$, R.~Legrain$^6$,
 N.~Le~Neindre$^5$, O.~Lopez$^5$, M.~Louvel$^5$,  A.M.~Maskay$^7$,
J.~Normand$^5$, P.~Paw{\l}owski$^1$, 
 E.~Rosato$^9$,  F.~Saint-Laurent$^3$,  J.C.~Steckmeyer$^5$, 
 B.~Tamain$^5$,  L.~Tassan-Got$^1$, E.~Vient$^5$, 
 J.P.~Wieleczko$^3$ \\
  INDRA Collaboration}

\address{
$^1$ Institut de Physique Nucl\'eaire, IN2P3-CNRS, F-91406 Orsay Cedex,
France.~\\
$^2$ National Institute for Physics and Nuclear Engineering, RO-76900
Bucharest-M\u{a}gurele, Romania.~\\
$^3$ GANIL, CEA et IN2P3-CNRS, B.P.~5027, F-14076 Caen Cedex, France.~\\
$^4$ Laboratorio Nazionale del Sud, Viale Andrea Doria, I-95129 Catania,
Italy.~\\
$^5$ LPC, IN2P3-CNRS, ISMRA et Universit\'e, F-14050 Caen Cedex, France.~\\
$^6$ DAPNIA/SPhN, CEA/Saclay, F-91191 Gif sur Yvette Cedex, France.~\\
$^7$ Institut de Physique Nucl\'eaire, IN2P3-CNRS et Universit\'e, F-69622 
Villeurbanne Cedex, France.\\
$^8$ Conservatoire National des Arts et M\'etiers, F75141 Paris cedex 03.\\
$^9$ Dipartimento di Scienze Fisiche e Sezione INFN, Universit\`a di Napoli
``Federico II'', I80126 Napoli, Italy.}

\date{\today}
\maketitle
\begin{abstract}
     Multifragmentation of a ``fused system'' was observed for central
     collisions between 32 MeV/nucleon $^{129}$Xe and $^{nat}$Sn. Most of the
     resulting charged products were well identified thanks to the high
     performances of the INDRA 4$\pi$ array. Experimental higher-order 
     charge correlations for fragments show a weak but non ambiguous
     enhancement of events with nearly equal-sized fragments. Supported by
     dynamical calculations in which spinodal decomposition is simulated, this
     observed
     enhancement is interpreted as a ``fossil'' signal of spinodal
     instabilities in finite nuclear systems.
\end{abstract}


The decay of highly excited nuclear systems through multifragmentation
 is, at present time, a subject of great interest in
nucleus-nucleus collisions, and many efforts are made to understand
the underlying physics. While this process has been observed for many 
years, its experimental knowledge  was strongly improved only recently 
with the advent of powerful $4\pi$ devices which authorize careful 
selections of well defined fused systems undergoing multifragmentation.
Many theories have been developed to explain multifragmentation 
(see for example ref.~\cite{MO93} for a general review of models).
One theory, in particular, is the concept of multifragmentation which
considers volume instabilities of the spinodal type.
Indeed, during a collision, a wide zone of the nuclear
matter phase diagram may be explored and the nuclear system may 
enter the liquid-gas phase coexistence region (at low density) and 
even more precisely the unstable spinodal region (domain of negative 
incompressibility)~\cite{BE83}. Thus, a possible origin of
multifragmentation may be found through the growth of density
fluctuations in this unstable region. Within this theoretical
scenario a
breakup into nearly equal-sized ``primitive'' fragments should be favored in
relation with the wave-lengths of the most unstable modes present in the 
spinodal region~\cite{AY95}. However this simple picture 
is expected to be strongly blurred by several effects:
the beating of different modes, eventual 
coalescence of nascent fragments, secondary decay of excited fragments
and mainly the finite size of the system~\cite{JA96}. Therefore only a weak
proportion of multifragmentation events with nearly equal-sized fragments is
expected. 
Note that surface instabilities developed
from peculiar multifragmenting geometrical structures (bubbles, disk-like),
 can also produce such events~\cite{MO93}.
In this letter we investigate, using a very sensitive method of charge
correlations, the occurrence
of equal-sized fragment partitions in a selected sample of experimental
events, corresponding to a  fused system undergoing
multifragmentation. A comparison is made with predictions of 3D
stochastic mean-field simulations of collisions which take into account 
the dynamics of the most unstable modes in the spinodal region.

\begin{figure}[htb]
\epsfxsize= 8.5cm
\epsffile{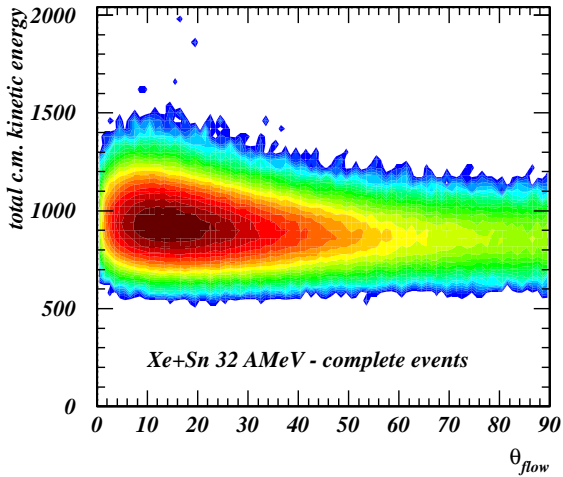}
\caption{Wilczynski diagram for complete events: correlation between total
 measured c.m. kinetic energy and flow angle $\theta_{flow}$. \label{bidim}}
\end{figure}

We have experimentally studied the reaction $^{129}Xe+^{nat}Sn$ at 32
MeV/nucleon. The measurements were performed at the GANIL accelerator using
the 4$\pi$ multidetector for charged reaction products
INDRA~\cite{indra}. Detailed information on the experiment and on the
selection of fused events can be found in Ref.~\cite{RI98,FR001}. Let
us just recall that the selection was performed by requiring the detection
of a significant fraction ($\geq$ 80\%) of the charge of the system; these
selected events are called complete events.
Reaction products with charge Z $\geq$ 5 were defined as fragments.
Finally the preferred direction of emission of matter in the
center of mass of the reaction (flow angle) was determined from the
calculation of the energy tensor of fragments, and the requirement was
made that this angle be larger than 60$^o$~\cite{LE96,FR001}. The main
argument underlying the chosen selection is that while a fused system
should be present at all flow angles, binary dissipative collisions
should vanish when this angle is large, giving way to an almost pure
``fusion'' phenomenon~\cite{BE96}.The present selection corresponds
to $\sim$25 mb, and the total cross-section for multifragmentation of
compact shape systems is estimated to 100 mb.
Figure~\ref{bidim} exhibits  how complete
 events populate the flow angle domain as a function of their measured
total kinetic energy (i.e. c.m. kinetic energy of all detected 
fragments and light charged particles).
An unambiguous charge identification is essential for building unbiased
charge correlation functions: in this experiment all fragments with 
charge Z$\leq$20 are identified with a resolution of one charge unit.
The correlation method used, called ``higher order correlations''
was proposed a few years ago in ref~\cite{MO96}. It is appropriate to 
search for weak signals and its originality consists of the fact that 
all information on fragments of one event is condensed in two variables 
to construct the charge correlation.
This quantity is defined by the expression:

\begin{equation}
\left. \frac{Y(\Delta Z, <Z>)}{Y'(\Delta Z, <Z>)} \right| _{M} 
\end{equation}

Here, $Y(\Delta Z, <Z>)$ is the yield of selected events with
$<Z>$ and $\Delta Z$ values;
$M$ is the fragment multiplicity, $<Z>$ denotes the average fragment charge
of the event:

\begin{equation}
<Z> = \frac{1}{M} \sum_{i=1}^{M} Z_i
\end{equation}

and $\Delta Z$ is the standard deviation, defined by:
\begin{equation}
\Delta Z = \sqrt{\frac{1}{M-1} \sum_{i=1}^{M} (Z_i - <Z>)^2}
\end{equation}

The denominator $Y'(\Delta Z, <Z>)$ which represents the  
uncorrelated yield is built, for each fragment multiplicity, by taking 
fragments in different events of the selected sample. The number of
uncorrelated events is
chosen large enough (10$^3$ per true event) to strongly reduce 
their contribution to statistical error. 
With such a correlation method, if events with  nearly equal-sized fragments
are produced, we expect to see peaks appearing in the first $\Delta$Z bin
(0-1).

\begin{figure}
\epsfxsize= 8.5cm
\epsffile{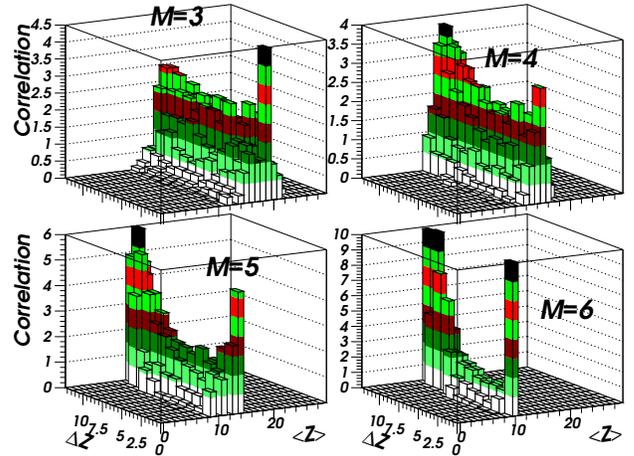}
\caption{Experimental higher-order charge correlations for selected fusion
events in 32 MeV/nucleon Xe+Sn collisions. } \label{cor_exp_SU}
\end{figure}

Higher order correlation functions for selected experimental events are
shown in fig.~\ref{cor_exp_SU}.
We observe peaks in the bin $\Delta$Z = 0-1 for each fragment 
multiplicity. From M=3 to M=6
the maximum is moving from $<Z>$=19 to $<Z>$=10. When multiplied by
the multiplicity these lead to a roughly constant value close
to the most probable total fragment 
charge for all fused events. It was moreover verified that the kinetic 
energy of fragments in these events  does not differ (within statistical 
error) from average kinetic energies of fragments for  other selected events.

\begin{figure}
\epsfxsize= 7.cm
\epsffile{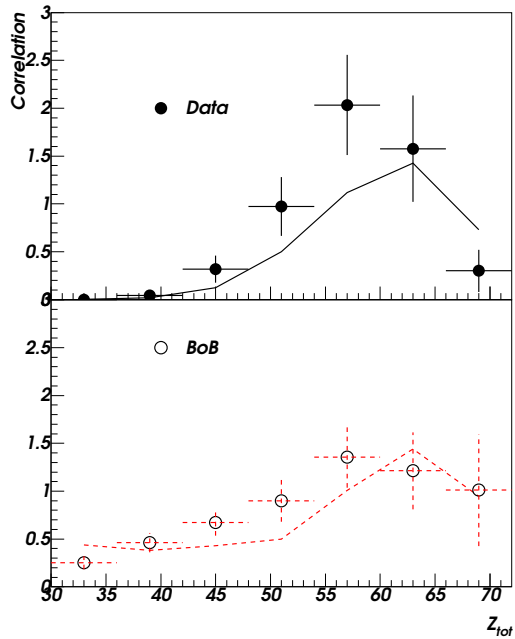}
\caption{Higher-order charge correlations: quantitative results
for experimental data (upper panel) and simulations (lower panel).
Symbols indicate the events where $\Delta$Z = 0-1, curves show
the background defined extrapolating $\Delta$Z $>$ 2 (see text).
Vertical bars correspond to statistical errors assuming independent
measurements and horizontal bars define $Z_{tot}$ bins.}
\label{cor_exp_quant}
\end{figure}

We can now estimate whether the enhancement of events with
equal-sized fragments is statistically significant and quantify their
occurrence. In this aim
we built charge correlations for all events, whatever
their multiplicity, by replacing the variable $<Z>$ by
$Z_{tot}$ = M$\times <Z>$. For this compact presentation uncorrelated events
are built and weighted
 in proportion to real events of each multiplicity.
 For each bin in $Z_{tot}$, fixed at six atomic number units, an exponential
 evolution of the correlation function is observed from $\Delta$Z=7-8 down to
 $\Delta$Z=2-3. This exponential evolution is thus taken as ``background'' to
 extrapolate down to the first $\Delta$Z bin.
  Higher order  correlation  functions for the first bin in $\Delta$Z
  are displayed in
 Fig.~\ref{cor_exp_quant} with their statistical errors;
 the solid line
 corresponds to the extrapolated ``background''. All events corresponding to
 the three points whose error bar is fully located above this
 line correspond to a statistically 
 significant enhancement of equal-sized fragment partitions. The
 probabilities that these values which are higher than the background 
 simply arise
 from statistical fluctuations are 0.048 ($Z_{tot}$=45), 0.038 ($Z_{tot}$=51)
 and 0.022 ($Z_{tot}$=57).
 The number of significant events amounts to 0.1\% of selected fusion events.

\begin{figure}
\epsfxsize= 8.5cm
\epsffile{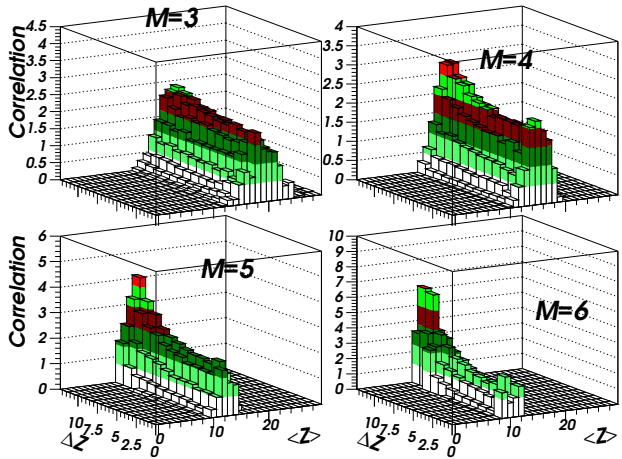}
\caption{Experimental higher-order charge correlations for very dissipative
``binary'' collisions in 32 MeV/nucleon Xe+Sn collisions (see text). } \label{cor_exp_tot}
\end{figure}

To test if these equal-sized fragment partitions were specific
- or not - of the selected fused events, we have studied a data
sample dominated by very dissipative collisions of binary type.
 This is done by selecting in the Wilczynsky diagram for
complete events (see fig.~\ref{bidim}) total c.m. energy
lower than 1200 MeV with a flow angle $\theta_{flow}$$\leq$ $30^{o}$.
Within this new selection we estimate that about 9$\times$10$^3$ fused 
events 
are mixed with about 3.6$\times$10$^5$ binary (or eventually strongly
deformed, or both) events.
The correlation functions are presented in fig.~\ref{cor_exp_tot}. 
No statistically significant signals are observed. The small peak
for M=6 corresponds to 0.0008\% of selected events. This is an indication
that only
fused events are characterized by an enhanced production of nearly
equal-sized fragments.

In ref.~\cite{RI98}, it was experimentally observed that charge
distributions for fusion/multifragmentation events were independent of 
the total mass of the incident nuclei (248 or 393 u), while fragment 
multiplicities scaled as their total charge. These properties were the 
first hint of a bulk effect for producing  fragments. 

Dynamical stochastic mean-field simulations~\cite{RA90,CHO91} were 
performed for head-on collisions only, thus neglecting shape effects.
Spinodal decomposition is simulated using 
the Brownian one-Body (BoB) dynamics~\cite{CH94,GU97,FR002}, which consists in
employing a Brownian force in the kinetic equations.
The ingredients of the simulations are as follows. The
self-consistent mean field potential~\cite{ZA73} chosen gives a 
soft equation of state (K$_{\infty}$= 200 MeV) and the
finite range of the nuclear interaction is taken into account using a
convolution with a gaussian function with a width of 0.9 fm~\cite{GUA96}. 
The addition of a term proportional to $\Delta \rho$  in the mean-field 
potential allows to well-reproduce the surface energy of ground-state 
nuclei~\cite{COL98}. This is essential in order to correctly describe the 
expansion dynamics of the fused system.  In the collision term a constant
$\sigma_{nn}$ value of 41 mb, without in-medium, energy, isospin or 
angle dependence is used~\cite{BER78}.
In the following step the spatial configuration of the primary fragments,
with their excitation energies as produced by BoB, was taken as input in the
SIMON code~\cite{ADN98} to follow the fragment deexcitation while preserving
space-time correlations. Finally the events were filtered
to account for the experimental device. These complete simulations
well reproduce multiplicity and charge distributions of fragments and 
their average kinetic energies~\cite{FR002}. 
To refine the comparison, higher-order charge correlations are 
calculated for the simulated events.
Although all events in the simulation arise from spinodal decomposition,
only a very small fraction, as suspected for finite systems~\cite{JA96},
exhibits final partitions with equal-sized
fragments. This fraction is not strongly modified by the de-excitation of
fragments; as an example Z=15 primary fragments produced in BoB
simulations (with their mass and excitation energy distributions) lead to a
secondary Z distribution centered at Z=14 with a standard deviation of 0.6.
Finally the proportion of statistically significant equal-sized
fragment partitions is similar (0.15\%) to the experimental one.
A detailed quantitative comparison is
displayed in fig.~\ref{cor_exp_quant}. The similarities between experimental
and calculated events allow to attribute all fusion-multifragmentation
events to spinodal decomposition. The peaks observed near $\Delta Z$=0 
in the higher-order charge correlations are thus fossil
fingerprints of the partitions expected from spinodal decomposition.
It was previously observed that multiplicity and Z distributions,
 and the average kinetic energy of fragments were also well
reproduced with the statistical model SMM~\cite{BON95,LEN99}. It is
interesting to underline that in this sample of SMM events no partition 
with $\Delta Z<$2 exists.
Note also that fragment velocity correlations~\cite{BT01} are
perfectly reproduced either in BoB simulations, or in the breakup of systems
with elongated shapes. As none of these configurations is strongly necked or 
close to a bubble,
we can exclude surface instabilities as a possible source of 
equal-sized fragments, in favor of bulk instabilities~\cite{RI98,BE00,XU00}.

In conclusion, we have investigated charge correlation functions for fusion
events which undergo multifragmentation. We have found
a weak but unambiguously enhanced production of  events with
equal-sized fragments. Supported by theoretical simulations we interpret
this enhancement as a signature of spinodal instabilities as the origin
of multifragmentation in the Fermi energy domain. 
Spinodal instabilities are thus shown for the first time in a finite 
system. Moreover the occurrence of spinodal decomposition shows
the presence of a liquid-gas coexistence region and 
gives a strong argument in favor of the existence of a first order 
liquid-gas phase transition in finite nuclear systems.

\end{document}